\documentclass[11pt]{article}
\pdfoutput=1

\usepackage[headsep=.2in]{geometry}
\usepackage[backend=biber,style=authoryear,
maxcitenames=2, maxbibnames=4,uniquelist=minyear,safeinputenc=true]{biblatex}
\usepackage{graphicx}
\usepackage{subcaption}
\usepackage[suffix=]{epstopdf}
\usepackage{hyperref}
\usepackage{authblk}
\usepackage{enumerate} 
\usepackage{amsmath}
\usepackage{ amssymb }
\usepackage{fancyhdr}
\title{THE EFFECT OF HETEROGENEITY ON FINANCIAL CONTAGION DUE TO OVERLAPPING PORTFOLIOS}
\date{\vspace{-5ex}}
\author[1]{OPEOLUWA BANWO\thanks{opeoluwa.banwo.13@ucl.ac.uk}}
\author[2,3]{FABIO CACCIOLI}
\author[2]{PAUL HARRALD}
\author[1]{FRANCESCA MEDDA}
\affil[1]{QASER Laboratory, University College London, Gower Street}
\affil[2]{Department of Computer Science, University College London, Gower Street}
\affil[3]{Systemic Risk Centre, London School of Economics and Political Sciences, London, UK}

\begin{document}
	\let\ref\autoref
	
	% Add the title section.
	\maketitle	
	
	\begin{abstract}
		We consider a model of financial contagion in a bipartite network of assets and banks recently introduced in the literature, and we study the effect of power law distributions of degree and balance-sheet size on the stability of the system. Relative to the benchmark case of banks with homogeneous degrees and balance-sheet sizes, we find that if banks have a power-law degree distribution the system becomes less robust with respect to the initial failure of a random bank, and that targeted shocks to the most specialised banks (i.e. banks with low degrees) or biggest banks increases the probability of observing a cascade of defaults. In contrast, we find that a power-law degree distribution for assets increases stability with respect to random shocks, but not with respect to targeted shocks. We also study how allocations of capital buffers between banks affects the system's stability, and we find that assigning capital to banks in relation to their level of diversification reduces the probability of observing cascades of defaults relative to size based allocations. Finally, we propose a non-capital based policy that improves the resilience of the system by introducing disassortative mixing between banks and assets.
	\end{abstract}

	\section{Introduction}
	Financial institutions are increasingly diversifying their balance sheet across several asset classes in order to reduce the idiosyncratic component of their portfolio risk. This has led to increased global connectivity in the portfolio holdings across several institutions \autocite{Battiston2012,JosselinGarnier2013}. However, recent studies including \autocite{Gai2010,May2010,Arinaminpathy2012,Caccioli2014,Caccioli2011,Nier2007} have shown that while increased interconnectivity can help diversify risk across the system, it also serves as a contagion propagating and amplification mechanism whenever a crisis is underway. This was partly the reason American International Group (AIG) was bailed out during the financial crisis as many of the biggest financial institutions had become exposed to it via derivative contracts \autocite[][provides more details]{Scott2012}. Financial institutions are connected directly via inter-institutional lending (e.g. interbank and repo transactions) and also indirectly through similar asset investments such as connections arising from overlapping portfolios. While the former has drawn the most attention from studies focusing on the role of counterparty and roll-over risks in propagating contagion \autocite{Caccioli2011,Gai2010,Gai2011,Arinaminpathy2012,May2010}, academics and policymakers have only recently begun paying close attention to the systemic risk posed by indirect connections associated with overlapping portfolios \autocite{Caccioli2014,Huang2013}. 
	
	These connections provide a contagion channel for the propagation of mark-to-market portfolio losses to one or more financial institutions due to depression in asset prices resulting from fire sales by a distressed institution holding the same assets. In some cases, these losses may be sufficient to cause additional institutions to become distressed thereby resulting in more rounds of asset fire sales and further depression in asset prices. The 2007 quant crisis, for instance, was caused by a similar scenario in which the fire sales liquidation of the portfolio of one equity hedge fund depressed prices of assets held by other funds causing them to embark on additional rounds of selling which depressed asset prices even further and resulted in large portfolio losses \autocite[see][for an elaborate discussion]{Khandani2007}. The existing literature on overlapping portfolios have only considered bank interlinkages arising from a single asset class \autocite{Cifuentes2005,Nier2007,Gai2010,Arinaminpathy2012}. However, \autocite{Caccioli2014} have recently generalised the fire sales model introduced in \autocite{Cifuentes2005} to the case of many assets. They characterised the stability of the financial system in terms of its structural properties including average degree, market crowding, leverage and market impact using a bipartite financial network model in which the contagion channel is formed through local portfolio overlaps between banks with homogeneous degrees. 
	
	However, their approach relies on the assumption of homogeneity in the degrees and sizes of all banks which may not necessarily be the case. In fact recent empirical studies \autocite{Guo2015,Braverman2014a,Marotta2015,DeMasi2012} show that real financial networks of common portfolio holdings and balance sheet size distributions deviate from this assumption. Specifically, they provide evidence of a power law in these distributions. Following this findings, we generalise the approach in \autocite{Caccioli2014} to account for power law in the degrees and sizes of banks. We refer to banks with low degrees as specialised while those with high degrees are said to be diversified. In this way, we are able to distinguish between the systemic risk contribution of different categories of banks ranging from very specialised to very diversified banks. Furthermore, we studied the effectiveness of various regulatory capital policy models guided by the intuition developed from the systemic risk contribution of the different types of banks. We then investigated the possibility of improving the system's stability by introducing structural correlation into the network without imposing new capital requirements. Finally, we characterise the stability response of the system with respect to leverage.
	
	The model used for our simulations belongs to the same class of contagion mechanisms used extensively in the literature of counterparty network models \autocite{Nier2007,Upper2011,Gai2010}. In a nutshell, the system is exogenously perturbed and the resulting impact is recursively propagated through the network until no new default is observed. This feedback mechanism is essentially driven by asset devaluations based on a market impact function that revalues an asset with respect to its traded volume \autocite{Bouchaud1998,Bouchaud2009}. Our goal is to understand the impact of heterogeneity in the portfolio structure of banks on financial contagion due to overlapping portfolios. As such, we abstract from strategic processes used by banks in choosing a particular portfolio structure as in \autocite{WAGNER2011}, who shows using a microfounded model that in equilibrium the risk of joint liquidation motivates investors towards heterogeneous portfolio configurations. Moreover, the mechanistic approach we consider keeps the model general enough for stress testing real financial systems by calibrating the model. We further assume passive portfolio management so as to keep the dynamics simple (i.e. banks do not deleverage or rebalance their portfolios during a crisis). In this sense, a bank's portfolio remains fixed until it becomes liquidated whenever it defaults. This assumption can be justified from the fact that most financial markets are illiquid relative to the \textbf{}positions held by large institutions such that whenever a crisis is underway, banks usually have insufficient time to deleverage until they become insolvent \autocite[see][for an elaborate discussion]{Caccioli2014}.
	
	Our stress tests reveal that heterogeneous bank degrees and sizes make the system more unstable relative to the homogeneous benchmark case with respect to random shocks but not with respect to targeted shocks. In contrast, heterogeneity in asset concentrations makes the system more resilient to random shocks but not with respect to targeted shocks. We then proceeded to study possible capital policy models guided by these results and find that a regulatory policy that assigns capital to the most specialised banks performs better than random assignments when the average degree is high. Moreover, diversification is a more significant factor than size in improving the financial system's resilience with capital based policies. The insights we develop can be used to address one of the major drawbacks of the Basel accords in ignoring the role of diversification for setting capital requirements \autocite{CEBS2010}. An example is the risk weighted capital requirement framework which is heavily criticised for providing banks with incentives to concentrate in low risk asset classes such as interbank loans, sovereign debt etc. which not surprisingly turned out to be at the centre of the 2007 financial crisis \autocite{Adrain2012}.  Finally, we investigated the possibility of improving financial stability with a non-capital based policy that imposes a particular configuration in the bipartite network and find that disassortative mixing (i.e. connecting the most specialised banks with the most concentrated assets) increases the stability of the system.
	
	The rest of this paper is organised as follows. In the next section, we outline the main features of the model. In \autoref{se3}, we explore the stability impact of heterogeneous network topology and balance sheet sizes. \autoref{se:pol} provides insights on the effectiveness of capital based policies and proposes a non-capital based policy by introducing structural correlations into the bipartite network. In \autoref{se:lev}, we study the impact of leverage on our model. Finally, a summary of our findings is presented in \autoref{se:con}.
	
	\section{The Model}
	\subsection{Network}
\begin{figure}[!htbp]
		\centering
		\includegraphics[width=.5\linewidth]{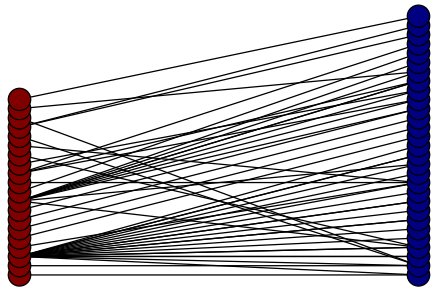}
		\caption{A Heterogeneous bipartite financial network. Banks are depicted in red circles while assets are shown in blue. The links of the banks follows a power law distribution}
		\label{fig0}
	\end{figure}

	As in \autocite{Caccioli2014}, we consider a bipartite network of a financial system consisting of $N$ banks and $M$ assets as shown in \autoref{fig0}. A link from bank $i$ to asset $j$ implies that $j$ constitutes part of the portfolio of bank $i$. We define $k_i$ as the degree (i.e. the total number of links) of bank $i$. Hence, the average bank degree is defined as:
	\begin{equation}
		\mu_b=\frac{1}{N}\sum_{i=1}^Nk_i
	\end{equation}
	Similarly, we can define the average degree of the assets as:
	\begin{equation}
		\mu_a=\frac{1}{M}\sum_{j=1}^Ml_i
	\end{equation}
	Where, $l_j$ is the number of banks holding asset $j$ in their portfolio. It is the case that the number of links emanating from both sides of the bipartite network must be equal i.e. $\mu_bN=\mu_aM$. Thus, we have that $\mu_b=\mu_a$ whenever $N=M$.
	\subsection{Balance sheet structure}
\begin{figure}[!htbp]
		\centering
		\includegraphics[width=.5\linewidth]{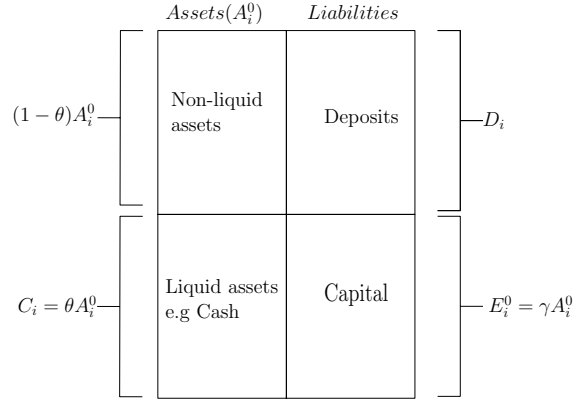}
		\caption{A typical bank's balance sheet structure. The bank holds a fixed amount of its asset in the form of cash, which value is assumed to remain fixed throughout the simulation for the purpose of simplicity.}
		\label{fig1}
	\end{figure}

	A typical bank's portfolio in the network discussed above consist of investments in non-liquid assets (e.g. shares in stocks) and liquid assets (e.g. cash). \autoref{fig1} depicts the general structure of a bank's balance sheet.
	We have defined a bank's proportion of liquid assets and initial capital as 20\% and 4\% of its total assets respectively for consistency with previous work \autocite{Caccioli2014,Gai2010}. Moreover, reports in \autocite{Upper2011} suggest that the capital structure of banks in advanced economies typically conforms with this configuration.
	We define the total asset of bank $i$ at any time $t$ is defined as:
	\begin{equation}
		A_i^t = \sum_{j=1}^MQ_{ij}p_j^t+C_i
	\end{equation}
	Where $Q_{ij}$ denotes the number of shares of stock $j$ held by bank $i$, $p_j^t$ is the price of stock $j$ at time $t$ defined as:
	\begin{equation}
		p_j^t=p_j^{t-1}f_j(x_j^t )
	\end{equation}
	Where $x_j^t$ denotes the quantity of asset $j$ sold at time $t$
	The capital (equity) of bank i at time t is given as:
	\begin{equation}
		E_i^t= A_i^t-D_i
		\label{eq5}
	\end{equation}
	In the model, a bank is declared insolvent whenever its initial capital endowment $E_i^0$ is completely eroded due to losses incurred from the depreciation of its asset values. Hence, we define the solvency condition for a bank $i$ as:
	\begin{equation}
		A_i^0-\sum_{j=1}^MQ_{ij}p_j^t-C_i\leq E_i^0
		\label{eq6}
	\end{equation}
	We can also express the solvency condition for bank i as a condition on its initial leverage defined as \(\lambda_i=A_i^0/E_i^0\) i.e.
	\begin{equation}
		\lambda_i\le \frac{\sum_{j=1}^MQ_{ij}p_j^t+C_i}{E_i^0}+1
		\label{eq7}
	\end{equation}
	Hence, leverage is a necessary condition for a bank to fail since an unleveraged bank i.e. $(\lambda_i=1)$ would always satisfy \autoref{eq6}.
	
	\subsection{Contagion mechanism}
\begin{figure}[!htbp]
		\centering
		\includegraphics[width=\linewidth]{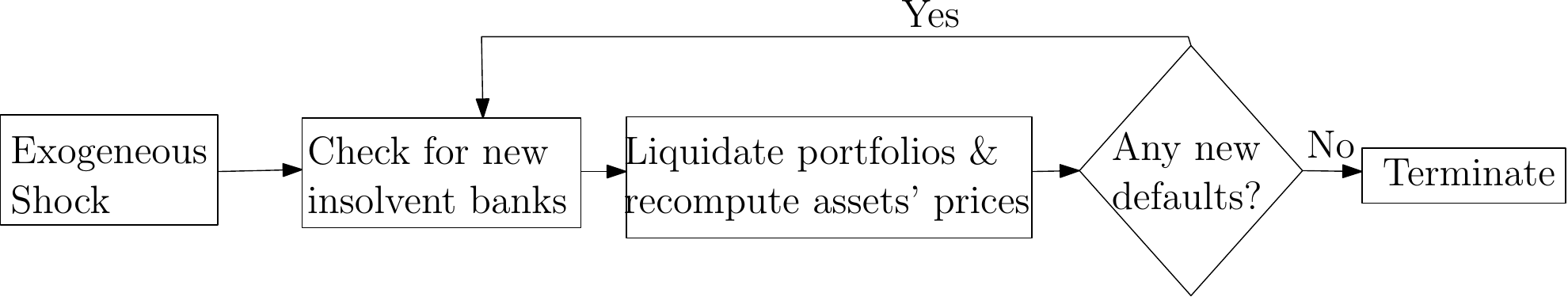}
		\caption{Flowchart representation of the contagion mechanism. A Bank is only declared bankrupt whenever it becomes insolvent.}
		\label{fig2} 
	\end{figure}

	A typical simulation in our model follows the sequence enumerated below:
	\begin{enumerate}[Step 1.]
		\item Exogenously shock the system at time step $t=0$
		\item Check banks for solvency condition as in \autoref{eq7} at each successive time steps $t=1,2,..$ 
		\item Liquidate the portfolios of any newly bankrupt banks and re-compute asset prices \footnote{In order to keep the model simple, we assume that the liquidated assets are traded with parties outside the banking system.}
		\item Terminate the simulation when no new default(s) occurs between successive time steps. 
	\end{enumerate}
	This dynamics is captured by the flowchart depicted in \autoref{fig2}
	\subsubsection{Exogenous shocks}
	We consider two kinds of initial shocks: random and targeted shocks. In a random shock, a bank or asset is randomly selected and exogenously perturbed while a specific kind of bank or asset is perturbed in the case of a targeted shock.
	\subsubsection{Market impact}
	We assume a market impact function of the form $f_j(x_j)=e^{-\alpha x_j}$ as in \autocite{Gai2010,Arinaminpathy2012,Cifuentes2005} such that $x_j$ is the liquidated fraction of asset $j$. The price of asset $j$ is then updated according to the rule: $p_j\rightarrow p_jf_j(x_j)$. As in \autocite{Gai2010,Nier2007,Caccioli2014}, we set $\alpha=1.0536$  such that the liquidation of 10\% of an asset results in a 10\% price drop in the asset's value.
	\subsubsection{Systemic stability}
	We characterise the stability of the financial system in terms of the systemic risk posed by an exogenous shock. We define systemic risk as the probability that contagion occurs. In the context of our model, contagion is said to occur only when the number of cascading defaults resulting from an exogenous shock exceeds a critical threshold $\phi$. We define $\phi$ as 5\% of the total number of banks in the system for consistency with previous work on financial contagion \autocite{Gai2010,Caccioli2014}.
	
	\section{Stability Analysis}
	\label{se3}
	The existing literature on financial contagion due to overlapping portfolios have only considered banks with homogeneous (i.e. similar) degrees and sizes \autocite[see][]{Caccioli2014,Cifuentes2005}, for instance, \autocite{Caccioli2014} consider a homogeneous financial network using an Erd\H{o}s-R\'enyi bipartite networks. However recent empirical studies by \autocite{Guo2015,Braverman2014a,Marotta2015,DeMasi2012} have shown that real portfolio networks are far removed from such distributions. In particular, they show the existence of a power law in the degree distributions in a network of overlapping portfolios similar to the observations reported in \autocite{Boss2004,Caccioli2015} for counterparty networks. 
	
	\subsection{Heterogeneous bank degrees}
	In this paper, we desire to investigate the stability impact of heterogeneity in the degree of banks. As such, we consider a heterogeneous bipartite financial networks where the degrees of banks are generated according to a power law distribution i.e. $P(k)\propto k^{-\gamma}$ with $\gamma=2.5$. Each bank then forms a link with a random asset until it reaches its generated degree such that no bank is linked to an asset more than once. This link formation approach implies that the number of links of the assets follows a Poisson distribution since every asset has the same probability of being selected. A bank's degree can be interpreted as its level of diversification since it denotes the number of different investments of the bank. We have used the term \textit{specialised bank} to mean a bank with focused investments in contrast to a bank holding a diversified portfolio. Our focus here lies in understanding the systemic risk contribution of different types of banks ranging from very specialised to very diversified banks without mixing in the influence of size. This approach mandates an assumption of the same balance sheet sizes across all banks. 
	
\begin{figure}[!htbp]
		\centering
		\begin{subfigure}{0.5\textwidth}
			\centering
			\includegraphics[width=.9\linewidth]{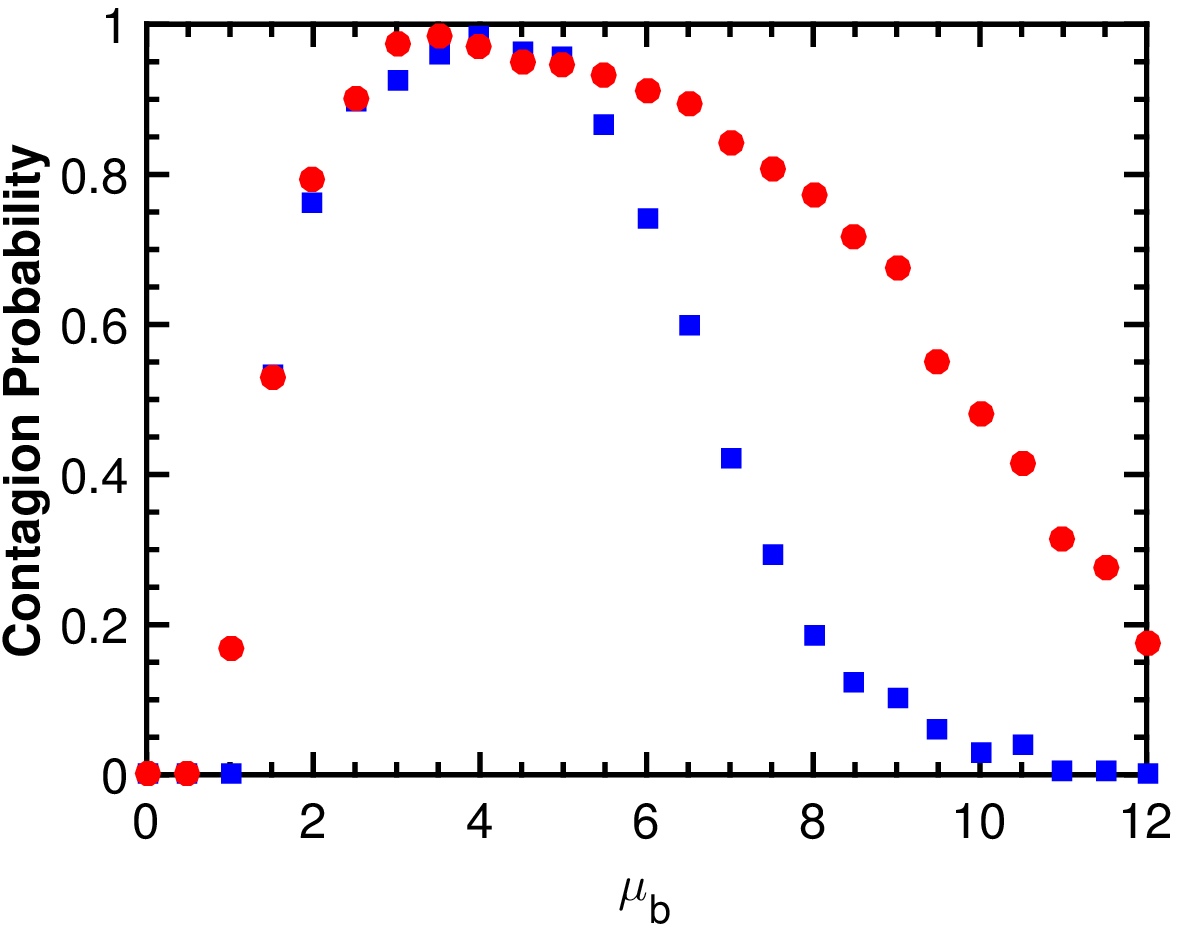}
			\caption{Bank shock}
			\label{f3sub1}
		\end{subfigure}%
		\begin{subfigure}{0.5\textwidth}
			\centering
			\includegraphics[width=.9\linewidth]{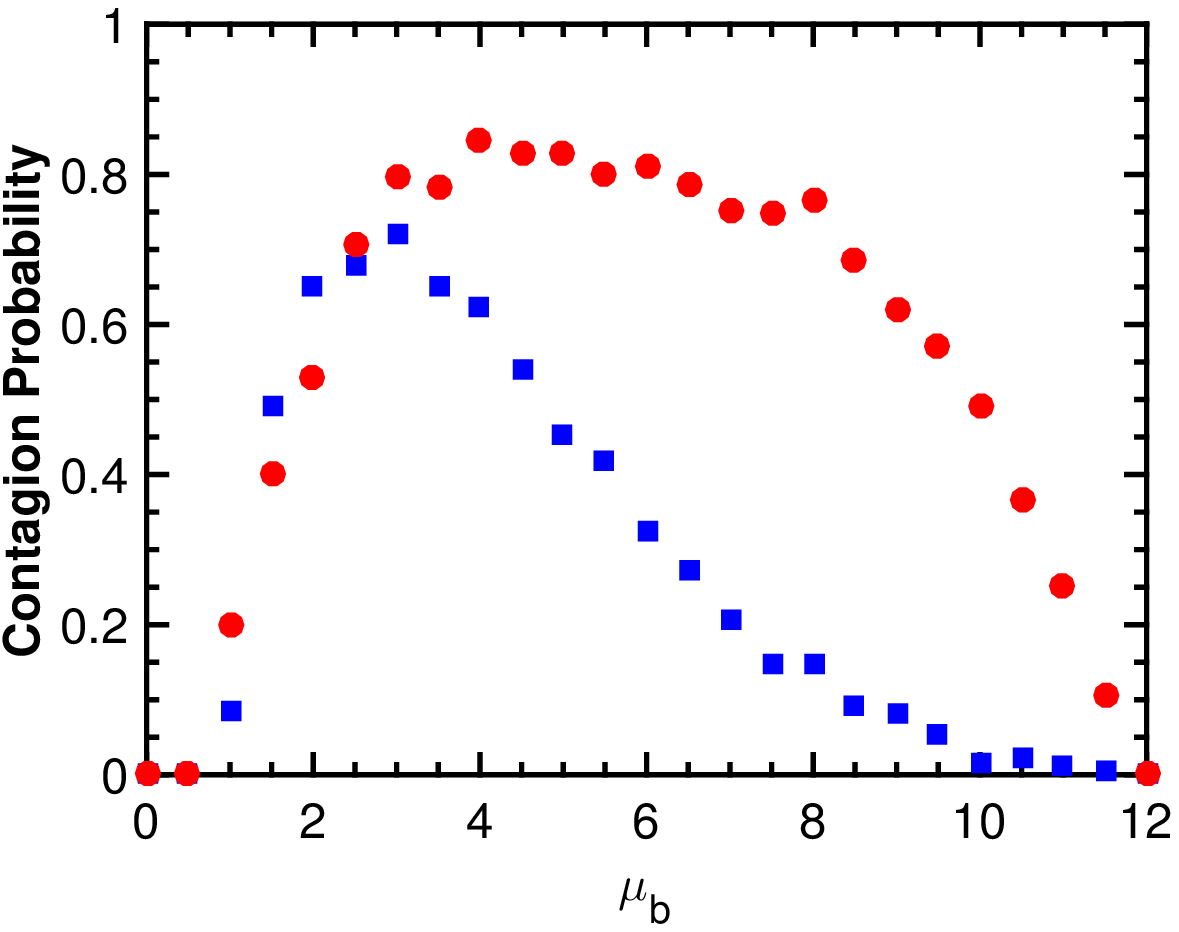}
			\caption{Asset shock}
			\label{f3sub2}
		\end{subfigure}
		\caption{Left Panel: Contagion probability as a function of $\mu_b$ for the case when a random bank fails. Red circles: system with heterogeneous bank degrees. Blue squares: system with homogeneous bank degrees. Right Panel: Contagion probability as a function of $\mu_b$ for the case when a random asset is devalued. Contagion is worse in the heterogeneous system irrespective of the kind of exogenous shock. Result refer to 1000 simulations for $N=M=1000$}
		\label{fig3}
	\end{figure}

	In the left panel of \autoref{fig3}, we plot the probability of contagion as a function of $\mu_b$ when a random bank fails. We compare the unstable region for the system with heterogeneous bank degrees relative to the homogeneous case. We find that the unstable region is wider in the heterogeneous system. The right panel of \autoref{fig3} shows that this observation is independent of the kind of exogenous shock. In particular, we plot the contagion probability for the case when an asset is randomly devalued and still find that heterogeneity in banks' degree results in greater instability.
	The existence of a wider unstable region in the heterogeneous system can be understood by observing that contrary to the homogeneous case, the heterogeneous system is characterized by a few highly diversified banks and many specialized banks. Hence, the probability that a specialized bank is hit from the initial shock is relatively higher. Consequently, specialised banks induce higher devaluations on their assets since they hold large amounts of these assets. 
	
	However, this result is in contrast to general reports in the complex networks literature in which heterogeneous network topology has been shown to create more stability, for instance, \autocite{Caccioli2011} show that heterogeneity in a counterparty network creates a more robust system relative to the homogeneous case. The reason for this lies in the fact these previous works have considered a network of direct bilateral exposures between the heterogeneous agents such that the few hubs (i.e. the most connected) nodes become the most systemically relevant whereas the specialised nodes are the most systemically relevant in this case due to the fact that they concentrate their investments in specific assets and thereby carry higher liquidation risk. This result sheds some light to why specialised institutions like mortgage banks, building and loan associations, specialist funds etc. who hold significant amounts of specific assets should be considered systemically important as the fire sales of these assets conditional on their default may have devastating impacts on asset prices. Moreover, this finding provides further credence to the conjecture given by Andrew Haldane, the Bank of England's Chief Economist, in one his speeches that the "rapid growth in specialist funds potentially carry risk implications, both for end-investors and for the financial system as a whole" \autocite{Haldane2014}. Furthermore, \autocite{WAGNER2011} also suggests imposing higher diversity requirements on portfolio holdings of financial institutions with high liquidation risk relative to those with low risk.
	
\begin{figure}[!htbp]
		\centering
		\includegraphics[width=.5\linewidth]{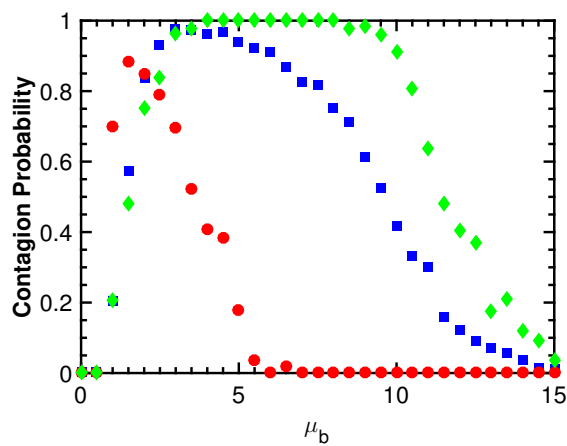}
		\caption{Contagion probability as a function of $\mu_b$ when banks have heterogeneous degrees. Blue squares: contagion probability when a random bank fails. Green diamonds: contagion probability when shocks are targeted at the most specialised banks. Red circles: contagion probability when shocks are targeted at only the most diversified banks. The region where contagion occurs is widest when specialised banks are targeted. Result refer to 1000 simulations for $N=M=1000$}
		\label{fig4}
	\end{figure}

	In \autoref{fig4}, we show the impact of targeted shocks on the stability of the system. We plot the probability of contagion as a function of $\mu_b$ when the initial shock is aimed at specific banks. We find that the unstable region is widest when any of the top 5\% most specialised banks is hit while targeted shocks on any of the top 5\% diversified banks results in the smallest unstable region. This can be understood from the fact that banks hold lesser amounts of specific assets with increasing degrees since we assume here that all banks are endowed with the same asset sizes. Hence, targeting shocks at the most diversified banks would effectively close the fire-sale contagion channel quicker since only small amounts of assets would be sold, which implies lower price devaluation than the case when banks are randomly perturbed. However, the reverse is observed when shocks are directed at the most specialised banks since they hold significant amounts of specific assets and thereby carry higher liquidation risk. We refer to these banks as "Too Specialised To Fail" (TSTF).
	
	\subsection{Heterogeneous asset concentration}
	In the previous section, we introduced heterogeneity into the distribution of the banks' degrees and the number of banks holding each asset is homogeneous. In this section, we turn our attention to the case when the distribution of the number of banks holding each asset class is heterogeneous and the degree distribution of banks is homogeneous. We follow the approach of the previous section and assume a power law distribution in the asset concentrations. An asset's concentration can be interpreted as the preference of banks towards that asset class. Our aim is to study how this preference structure affects the stability of the entire system.
	
	\begin{figure}[!htbp]
		\centering
		\begin{subfigure}{0.5\textwidth}
			\centering
			\includegraphics[width=.9\linewidth]{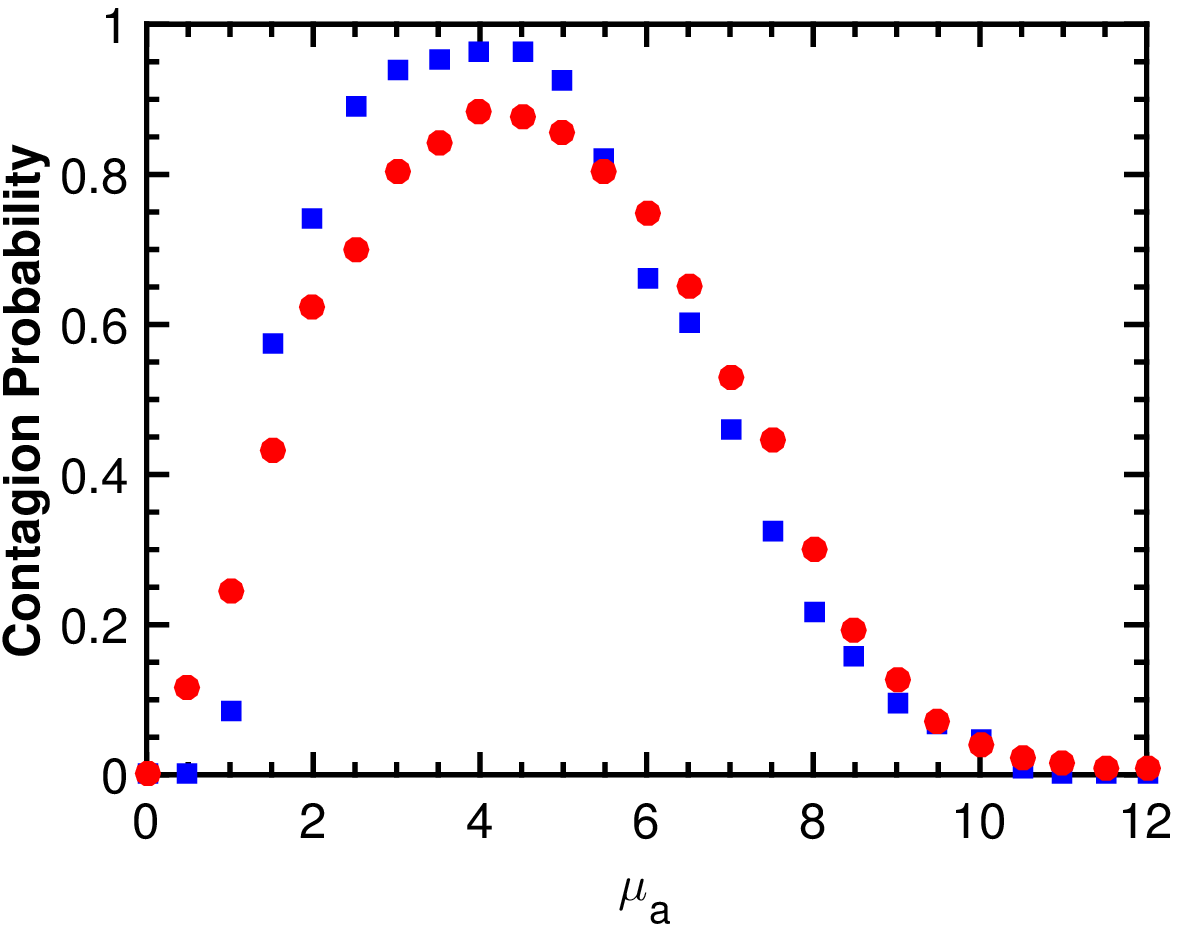}
			\caption{Random shocks}
			\label{f5sub1}
		\end{subfigure}%
		\begin{subfigure}{0.5\textwidth}
			\centering
			\includegraphics[width=.9\linewidth]{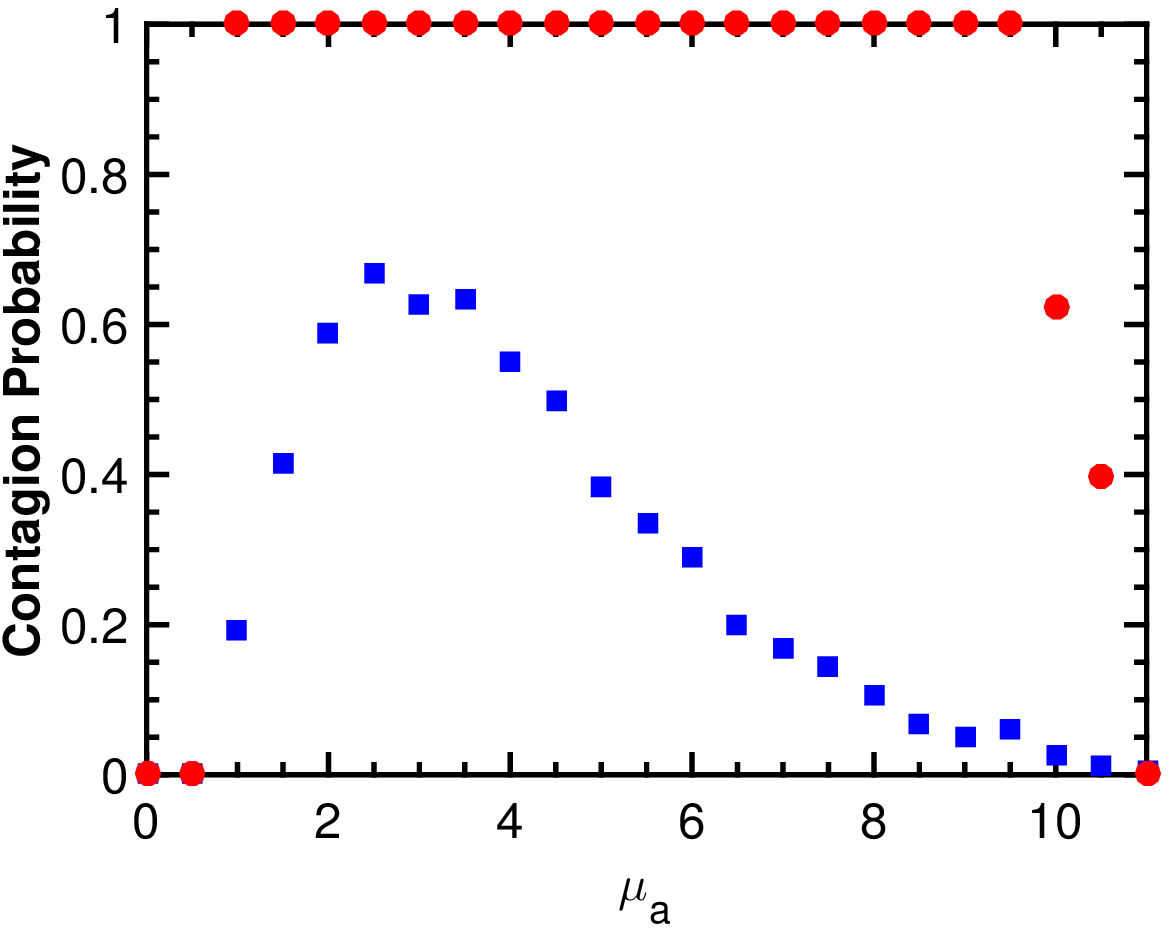}
			\caption{Targeted shocks}
			\label{f5sub2}
		\end{subfigure}
		\caption{Left Panel: Contagion probability as a function of $\mu_a$ for homogeneous and heterogeneous distributions of asset concentrations. Blue squares: system with homogeneous asset concentrations. Red circles: system with heterogeneous asset concentrations. A random bank fails in both cases. Introducing heterogeneity into the distribution of asset concentrations results in a more robust system. Right Panel: Targeted shocks on a system with heterogeneous asset concentrations. Targeting concentrated assets amplifies contagion probability. Result refer to 1000 simulations for $N=M=1000$}
		\label{fig5}
	\end{figure}
	
	In the left panel of \autoref{fig5}, we plot the probability of contagion as a function of average asset concentration for the case when a random bank fails. In contrast to the results observed for heterogeneous bank degrees, we find that introducing heterogeneity in the concentration of the assets produces a more robust system relative to the homogeneous system. This can be understood from the fact that the probability than a highly concentrated asset is perturbed is relatively low since the scale free network comprises very few concentrated assets and many less concentrated (i.e. isolated) ones. This effectively reduces the unstable region since fewer banks are affected by contagion. 
	
	The right panel of \autoref{fig5} shows the stability impact of aiming initial shocks at any of the top 5\% most concentrated assets. As expected, targeting initial shocks at these highly concentrated assets has the effect of amplifying contagion since more banks' portfolios are negatively affected by the initial asset devaluation. However, the width of the unstable region is essentially the same as in the homogeneous system. This is so because as soon as banks reach a critical average degree they become resilient to contagion irrespective of the kind of shock on the asset side. 
	
	\subsection{Heterogeneous bank sizes}
	In the previous sections, we assumed that all banks have the same balance sheet sizes in order to separate the influence of size from diversification. However, empirical evidence in the literature clearly suggest that banks also have largely heterogeneous sizes \autocite{Boss2004}. For instance, a recent data analysis by SNL Financial shows that the top 5 biggest banks have 44\% of the total assets held by banks in the U.S. \autocite{Schaefer2014}. Our aim in this section is to study the impact of this kind of heterogeneity in the size distribution of banks on the stability of the financial system. To do this, we model the bank sizes according to a power law distribution i.e. $P(A)\propto A^{-\gamma}$ resulting in the creation of a few banks with significantly larger asset sizes than most banks whilst abstracting from the influence of diversification by assuming a Poisson degree distribution.
	
	In the left panel of \autoref{fig7}, we plot the probability of contagion as a function of $\mu_b$ for the case of random bank shocks. We find that contagion halts much faster when banks have homogeneous sizes relative to the heterogeneous case. The following argument provides an intuition to why this is the case. In the heterogeneous system, the fire sales impact on asset prices is more severe whenever any of the large banks are hit as these banks hold significant amounts of their assets relative to the entire system since we have assumed a Poisson degree distribution. This effectively shifts the critical threshold for which contagion is no longer possible to the right. 
	
	\begin{figure}[!htbp]
		\centering
		\begin{subfigure}{0.5\textwidth}
			\centering
			\includegraphics[width=.9\linewidth]{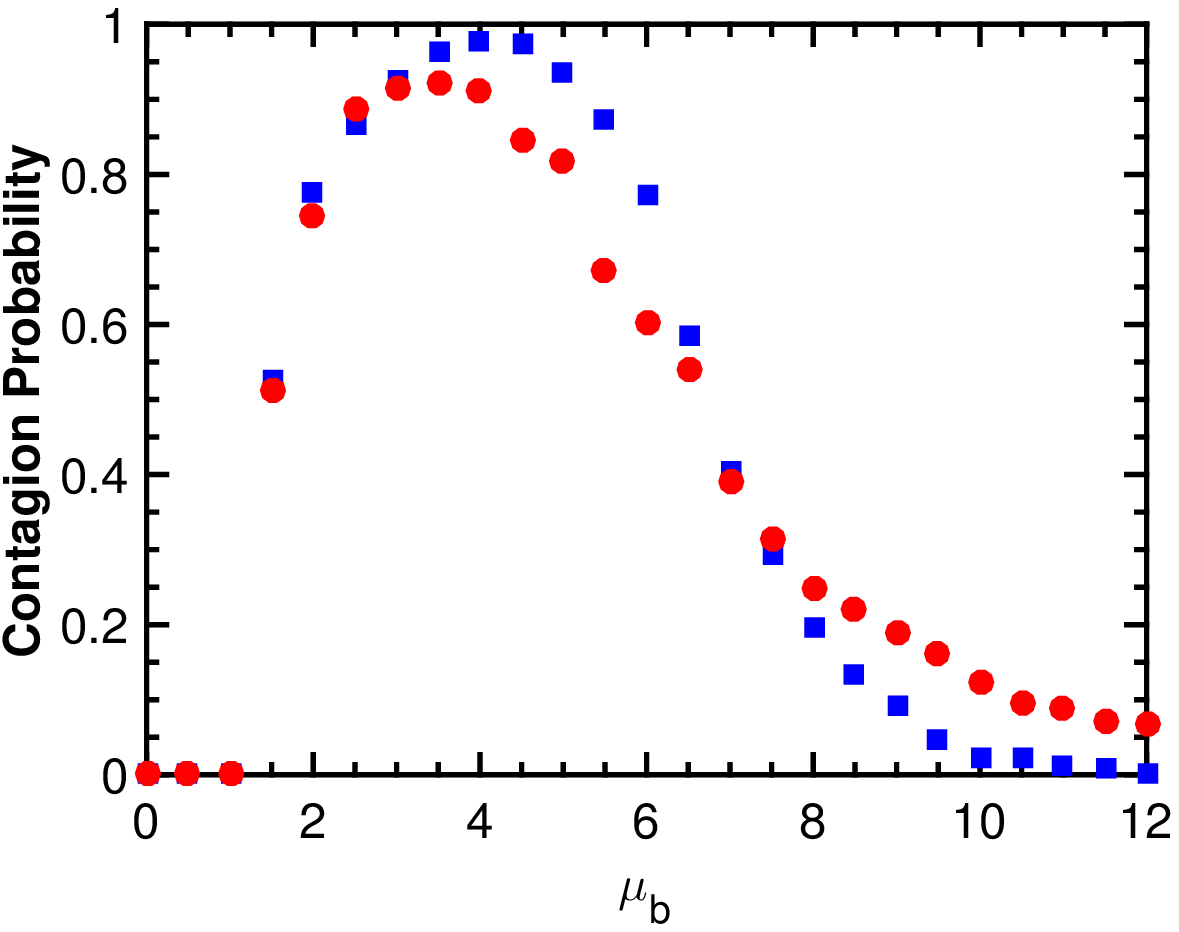}
			\caption{Random shocks}
			\label{f6sub1}
		\end{subfigure}%
		\begin{subfigure}{0.5\textwidth}
			\centering
			\includegraphics[width=.9\linewidth]{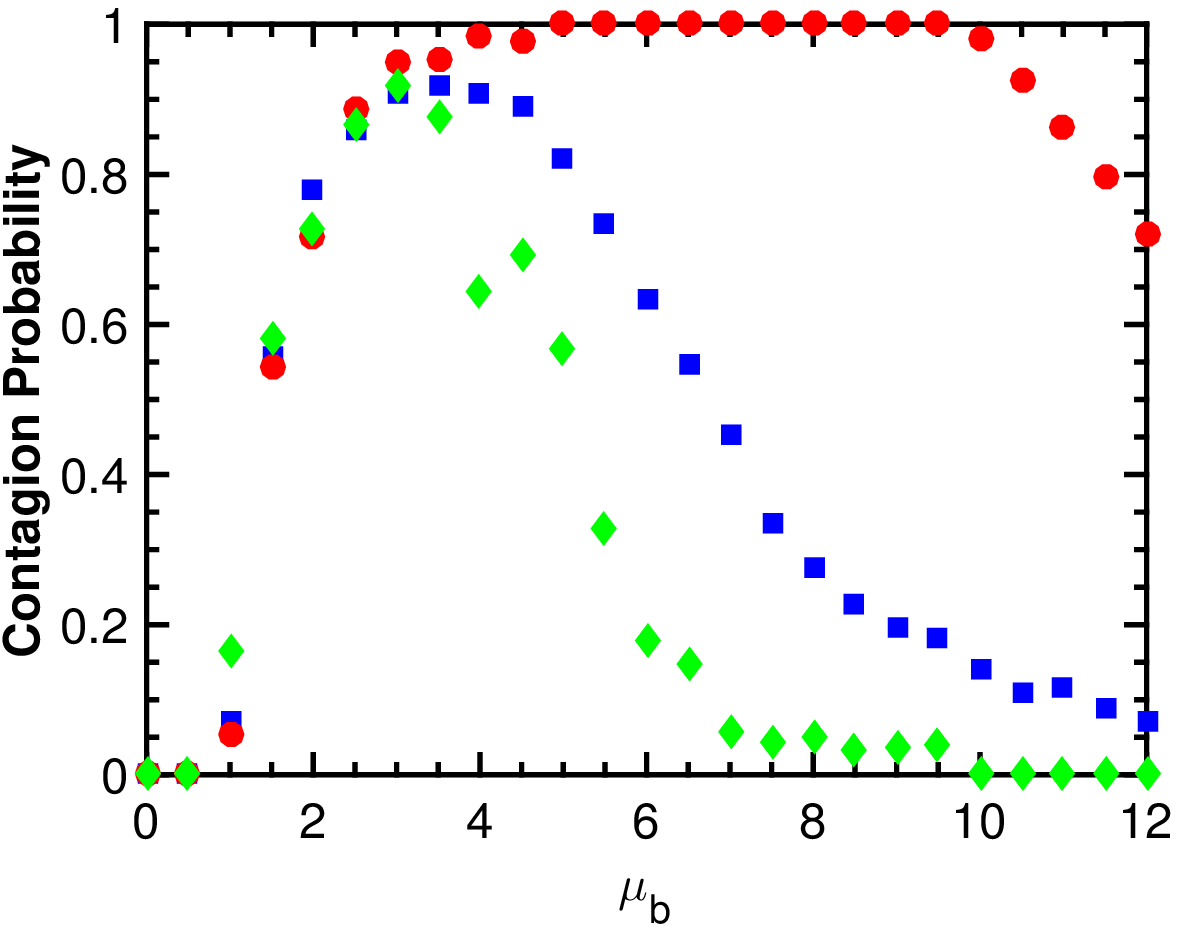}
			\caption{Targeted shocks}
			\label{f6sub2}
		\end{subfigure}
		\caption{Left Panel: contagion probability as a function of $\mu_b$ for homogeneous and heterogeneous distribution of banks' sizes. Blue squares: system with similar balance sheet sizes. Red circles: system with heterogeneous balance sheet sizes. The system is subject to random bank failures in both cases. Contagion probability is wider in the heterogeneous system relative to the homogeneous case. Right Panel: Targeted shocks on a system with heterogeneous distribution of banks' balance sheet sizes. Blue squares: contagion probability when a random bank is perturbed. Red circles: contagion probability when shocks are targeted at the biggest banks. Green diamonds: contagion probability when shocks are targeted at the smallest banks. Targeting shocks at the biggest bank results in the widest unstable region. Result refer to 1000 simulations for $N=M=1000$.}
		\label{fig6}
	\end{figure}
	
	The right panel shows the contagion probability as a function of $\mu_b$ for the case of initial shocks to specific banks. We observe that the system is significantly more unstable when exogenous shocks are targeted at any of the top 5\% biggest banks but more stable when the shocks are targeted at any of the top 5\% smallest banks. This follows from the fact that big banks hold comparatively larger amounts assets for each value of $\mu_b$ relative to other banks, which implies that targeting shocks at them would cause higher devaluations of the asset classes they hold, effectively fuelling the contagion mechanism that leads to a wider unstable region. We refer to these banks as "Too Big To Fail" (TBTF).  
	
	In summary, the findings of the stress tests conducted in \autoref{se3} are the following:
	\begin{enumerate}[(i)]
		\item Introducing heterogeneity in the degrees of banks exacerbates the fragility of the system to random shocks in contrast to \autocite{Caccioli2011,Gai2010} who show that a scalefree counterparty network results in a more robust system with respect to random shocks. We find that this result is independent of the type of exogenous shock (i.e. bank or asset shock). Furthermore, we find that targeting the most specialised banks makes the system more unstable. 
		\item Heterogeneity in asset concentrations improves the resilience of the system to random shocks in contrast to heterogeneous bank degrees. Moreover, targeting highly concentrated assets increases the probability of contagion, however the average degree threshold where contagion dies out is effectively unchanged.
		\item Cascading default is halted slightly faster when banks have homogeneous sizes relative to the heterogeneous case and is greater when exogenous shocks are targeted at the biggest banks.
	\end{enumerate}
	
	\section{Policy Impact Analysis}
	\label{se:pol}
	The 2007-2009 financial crisis has precipitated calls for higher regulatory capital requirements for banks.  Although higher capital requirements can improve financial stability, they however carry some implicit costs \footnote{This is based on the assumption that Modigliani-Miller theorem does not hold, which essentially implies that  a bank's capital structure does not affect profit or social welfare in an idealised world without frictions such as interest payments on debts, taxes, bankruptcy and agency costs \autocite{Miller1958}.} namely reduced profitability for banks and higher lending cost which may have a negative impact on social welfare \autocite{IMF2016,Bridges2014,Brooke2015}. Hence, it is important that new regulatory capital requirements are assigned to banks in the way that gives the most stable configuration. To this end, we investigate how the intuition developed from the stress tests in \autoref{se3} can influence capital based regulatory policies. We then propose an alternative non-capital based policy by studying the structure of the bipartite network.
	
	\subsection{Capital based policy}
	Here, we compare the performance of possible capital policy models following the intuition developed in \autoref{se3}. In each model, the same amount of capital $\chi$ is injected into the system. The difference in the policies lies in the way $\chi$ is distributed amongst the banks.
	
	\subsubsection{Targeted versus random}
	The stress tests done in \autoref{se3} suggests that the "Too Specialised To Fail" and "Too Big To Fail" banks are systemically important. Hence, it becomes interesting to ask if assigning capital requirements to only this group of banks can improve financial stability relative to targeting a random group of banks. We consider two kinds of targeted policies. In one, we assign the capital equally to only the top 5\% most specialised banks and refer to this policy as $\boldsymbol{T_S}$ while in the second, which we call $\boldsymbol{T_B}$, only the top 5\% biggest banks are required to hold more capital. We model a random policy for the purpose of comparison. In the random policy, 5\% of the banks are randomly selected and assigned additional capital requirements equally.
	
\begin{figure}[!htbp]
		\centering
		\begin{subfigure}{0.5\textwidth}
			\centering
			\includegraphics[width=.9\linewidth]{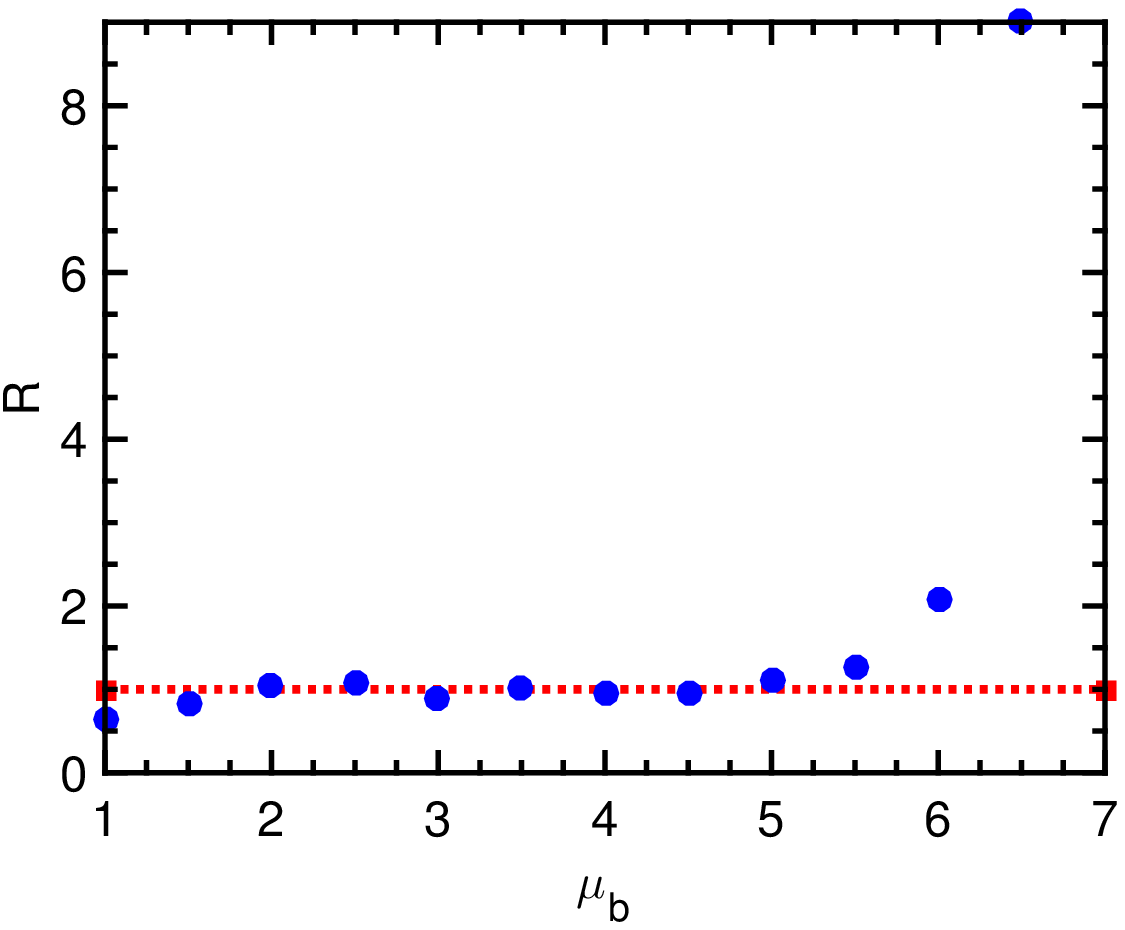}
			\label{f7sub1}
		\end{subfigure}%
		\begin{subfigure}{0.5\textwidth}
			\centering
			\includegraphics[width=.9\linewidth]{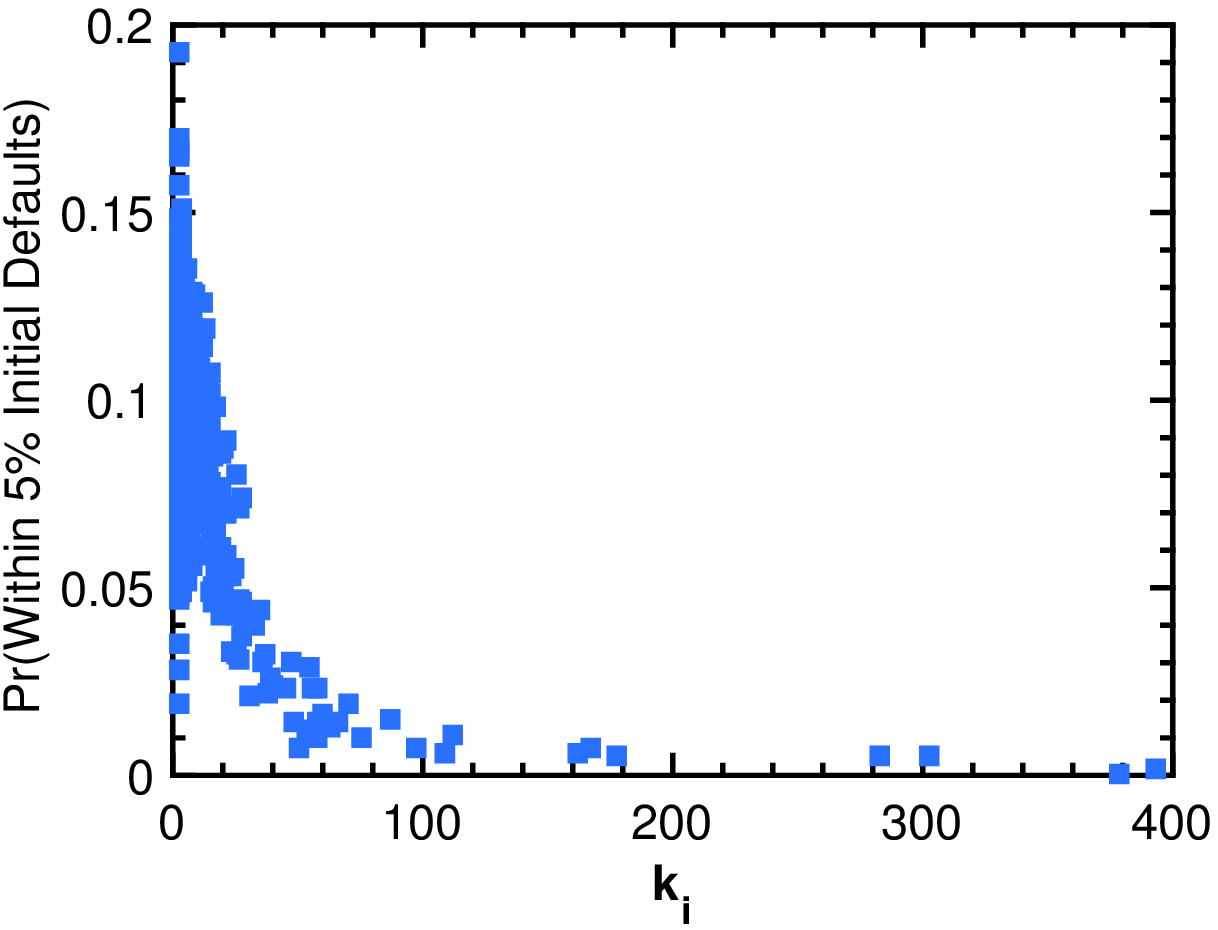}
			\label{f7sub2}
		\end{subfigure}
		\caption{Left panel: Stability impact of $T_S$ policy relative to the random policy for a system with heterogeneous bank degrees. Dotted line: comparison basis i.e. R=1. The $T_S$ policy produces more stability relative to the random policy for high values of $\mu_b$. Right panel: Probability that a bank i with degree $\mu_i$ defaults before contagion occurs. The most specialised banks have a greater chance of defaulting before contagion occurs.}
		\label{fig7}
	\end{figure}

	\paragraph{$\boldsymbol{T_S :}$}
	We now investigate the stability impact of the $T_S$ policy relative to the random policy as such we abstract away from the influence of size by assuming similar balance sheet sizes across all banks. We show this comparison in left panel of \autoref{fig7} by computing the ratio $R$ of the contagion probability of both policies as a function of $\mu_b$ such that $R=1$ implies similar performance, $R>1$ means the $T_S$ policy supersedes the random policy and $R<1$ implies that the $T_S$ policy outperforms the random policy. We focus our analysis on only those regions where contagion occurs in both systems to avoid divisions by zero. The plot suggests that a policy that focuses on the most specialised banks results in greater stability relative to a random policy in the region with high values of $\mu_b$, which is significant from a policy perspective because real world financial networks are more likely to be in this region.
	
	The right panel of \autoref{fig7} provides an insight to why the $T_S$ policy outperforms the random policy. It shows the probability that a bank $i$ with degree $k_i$ defaults before the occurrence of contagion. The plot suggest that the specialised banks are the most likely to default before contagion occurs. As such, it is reasonable to conjecture that focusing the capital policy on these banks is more likely to increase the resilience of the system.
	
\begin{figure}[!htbp]
		\centering
		\begin{subfigure}{0.5\textwidth}
			\centering
			\includegraphics[width=.9\linewidth]{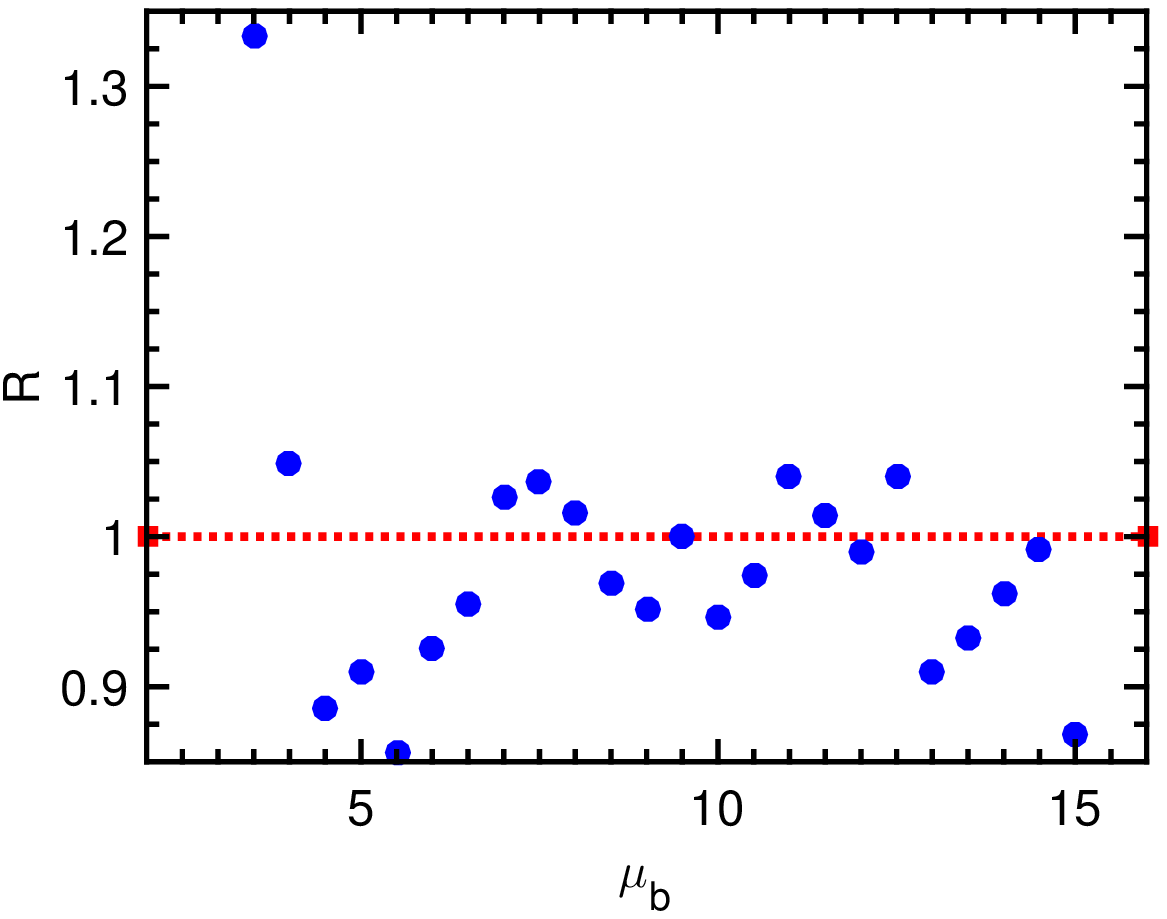}
			\label{f8sub1}
		\end{subfigure}%
		\begin{subfigure}{0.5\textwidth}
			\centering
			\includegraphics[width=.9\linewidth]{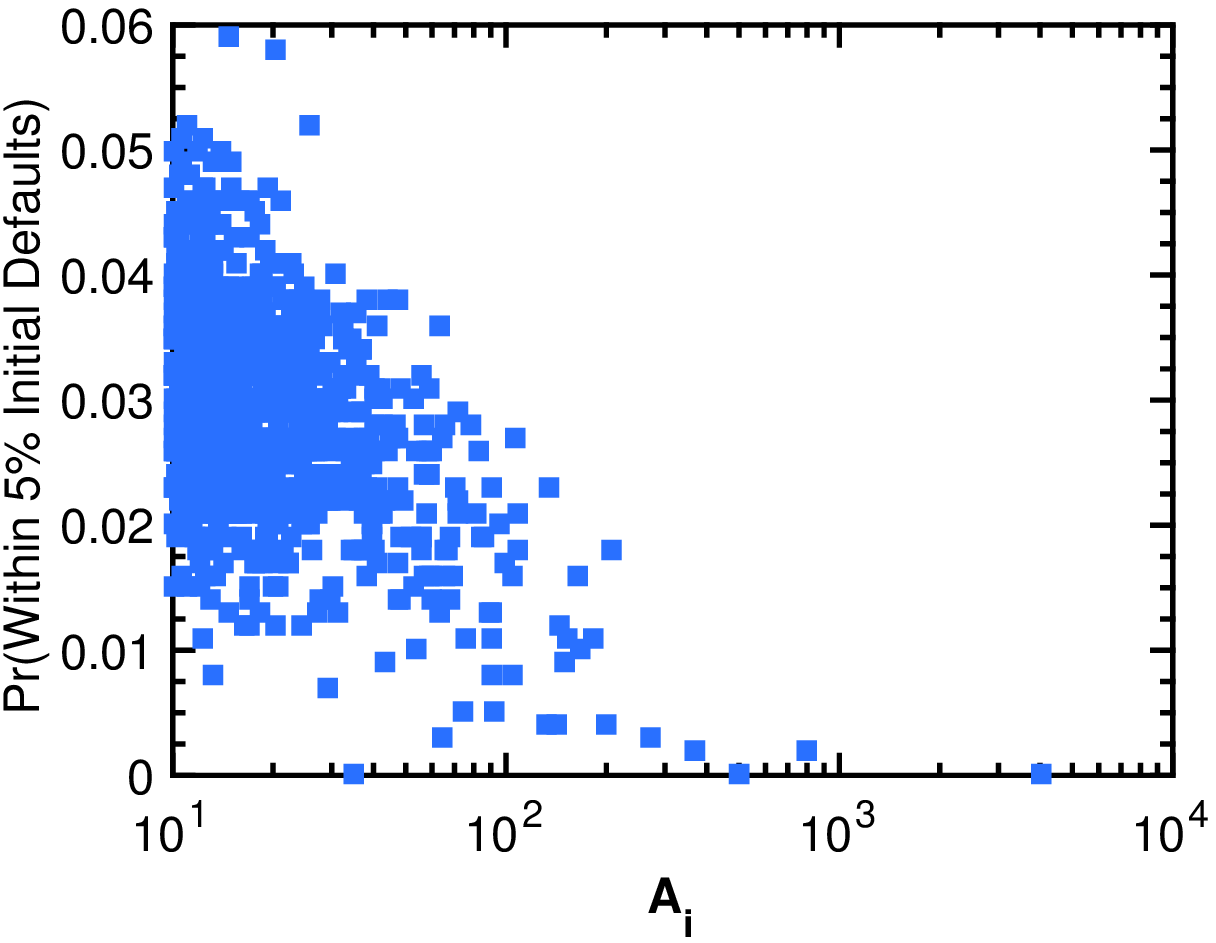}
			\label{f8sub2}
		\end{subfigure}
		\caption{Left panel: Stability impact of $T_B$ policy relative to the random policy for a system with heterogeneous bank sizes. Dotted line: comparison basis i.e. $R=1$. The $T_B$ policy appears to be ineffective relative to the random policy. Right panel: Probability that a bank $i$ with size $A_i$ (shown in log scale) defaults before contagion occurs. The biggest banks have a greater chance of defaulting before the occurrence of contagion.}
		\label{fig8}
	\end{figure}

	\paragraph{$\boldsymbol{T_B :}$}
	We now abstract from heterogeneous degrees and consider only heterogeneous sizes in order to study the stability impact of the $T_B$ policy relative to the random policy. We show this comparison in left panel of \autoref{fig8} by computing the ratio $R$ of the contagion probability of both policies as a function of $\mu_b$ such that $R=1$ implies similar performance, $R>1$ means the $T_B$ policy supersedes the random policy and $R<1$ implies that the $T_B$ policy outperforms the random policy. The plot markers oscillate around 1 suggesting that a policy that focuses only on the biggest banks is not effective.
	
	In order to understand why the $T_B$ policy does not perform better than the random policy, we plot the probability that a bank $i$ with size $A_i$ defaults before the occurrence of contagion in the right panel of \autoref{fig8} and find that big banks have a smaller chance of failing before contagion occurs. This implies that allocating capital requirements to only these banks is likely to be ineffective in the context of this model.
	
	\subsubsection{Diversification versus size}
	In the previous section, we simplified the model in order to separate the impact of diversification and size. However, it is also interesting to ask which of the two factors namely diversification and size is the more significant factor for capital requirement policies. In order to facilitate this comparison, we introduce heterogeneity into the degrees and sizes of the banks. The diversification based policy we consider assigns capital requirements to banks based on their degrees such that banks with higher degrees are required to hold lesser capital i.e.
	\begin{equation}
		\epsilon_i=\frac{1/k_i}{\sum_i 1/k_i}\chi
	\end{equation}
	Where, $k_i$ denotes the degree of bank $i$. While the size based policy allocates capital requirements to banks based on the size of their balance sheets such that big banks are required to hold more capital i.e.
	\begin{equation}
		\epsilon_i=\frac{A_i}{\sum_i A_i}\chi
	\end{equation}
	
\begin{figure}[!htbp]
		\centering
		\includegraphics[width=.5\linewidth]{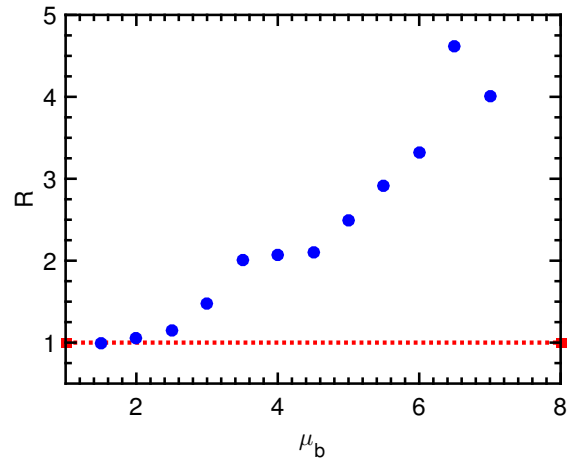}
		\caption{Stability impact of policy based on diversification relative to policy based on size as a function of $\mu_b$ for a system with heterogeneous sizes and degrees. Using banks' diversification levels as a proxy for assigning capital requirements is superior to using bank sizes.}
		\label{fig9}
	\end{figure}

	Where, $A_i$ denotes the size of bank $i$. In \autoref{fig9}, we compare the stability impact of a diversification based policy relative to a size based policy by computing the ratio $R$ of their respective contagion probabilities as a function of $\mu_b$ such that $R=1$ implies similar performance, $R>1$ means the diversification based policy supersedes the size based policy and $R<1$ implies that the size based policy outperforms the diversification based policy. The figure suggests that assigning capital based on a bank's degree supersedes assignment based on size further confirming recent findings reported by \cite{Cai2012}.
	
	\subsection{Non-capital based policy}
	\label{se:corr}
	From a policy maker's perspective, it is interesting to ask if there is a network structure that improves systemic stability without imposing new capital requirements \autocite[see][for example]{Thurner2013}? We address this question by introducing some structural correlation into the bipartite network. In the subsequent paragraphs, we use the term "assortative network" for a bipartite network in which the most diversified banks hold the most widely held (i.e. concentrated) assets and "disassortative network" for one in which the most specialised banks hold the most widely held assets while the most diversified banks hold the least held assets. 
	The correlated networks are generated based on the algorithm proposed in \cite{Noh2007}. The procedure essentially involves minimising a network cost function until a stationary state using Monte Carlos simulations. This cost function is defined as:
	\begin{equation}
		H(G)=-\frac{J}{2}\sum_{i,j=1}^Na_{ij}k_ik_j
	\end{equation}
	\[
	a_{ij}
	\begin{cases}
	0, & \text{if}\ i=j \\
	1, & \text{otherwise}
	\end{cases}
	\]
	Where, $k_i=\sum_ja_{ij}$  and $J$ denotes a control parameter for tuning the level of assortativity i.e. $J<0 (J>0)$ gives a disassortative (assortative) network respectively while $J=0$ produces an uncorrelated network. 
	\begin{figure}[!htbp]
		\centering
		\begin{subfigure}{0.5\textwidth}
			\centering
			\includegraphics[width=.9\linewidth]{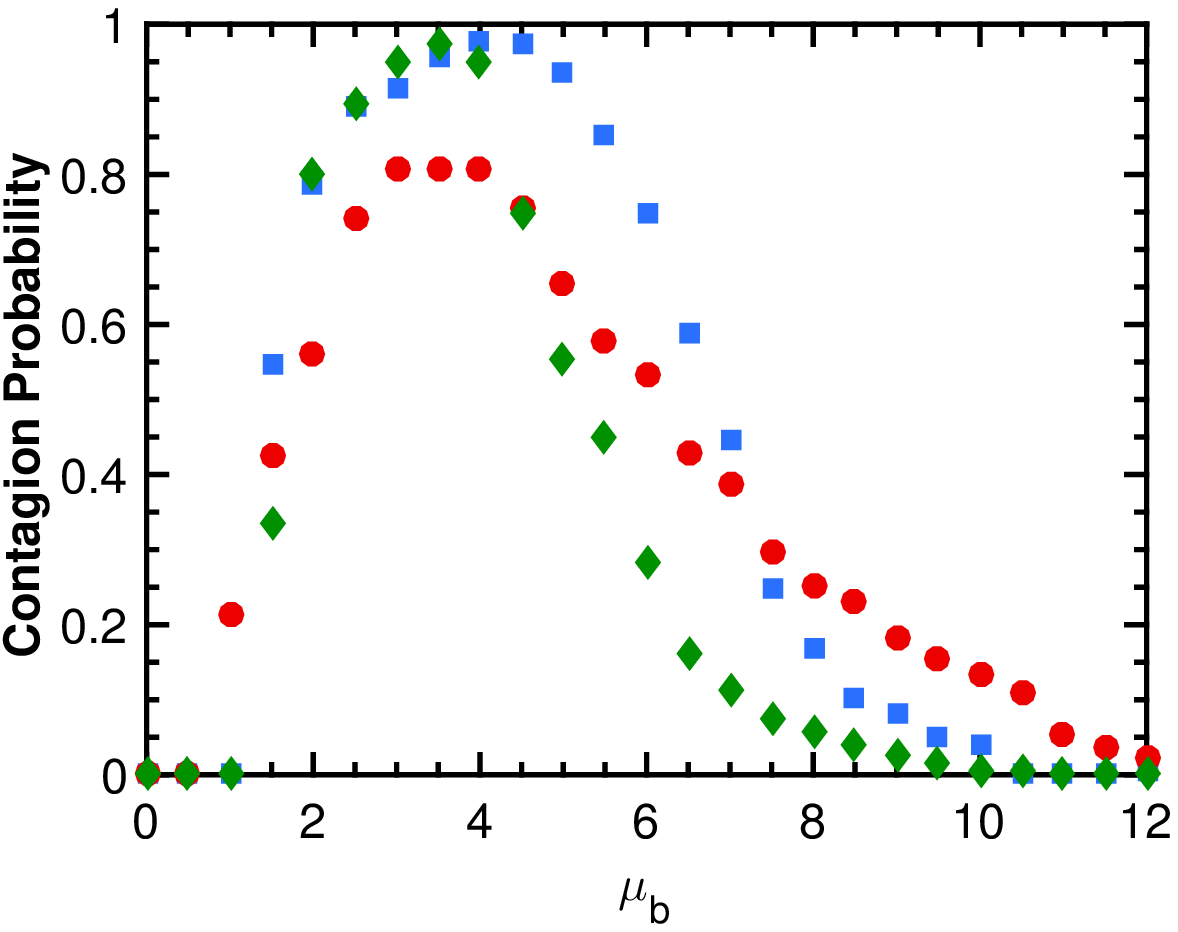}
			\caption{Bank shock}
			\label{f10sub1}
		\end{subfigure}%
		\begin{subfigure}{0.5\textwidth}
			\centering
			\includegraphics[width=.9\linewidth]{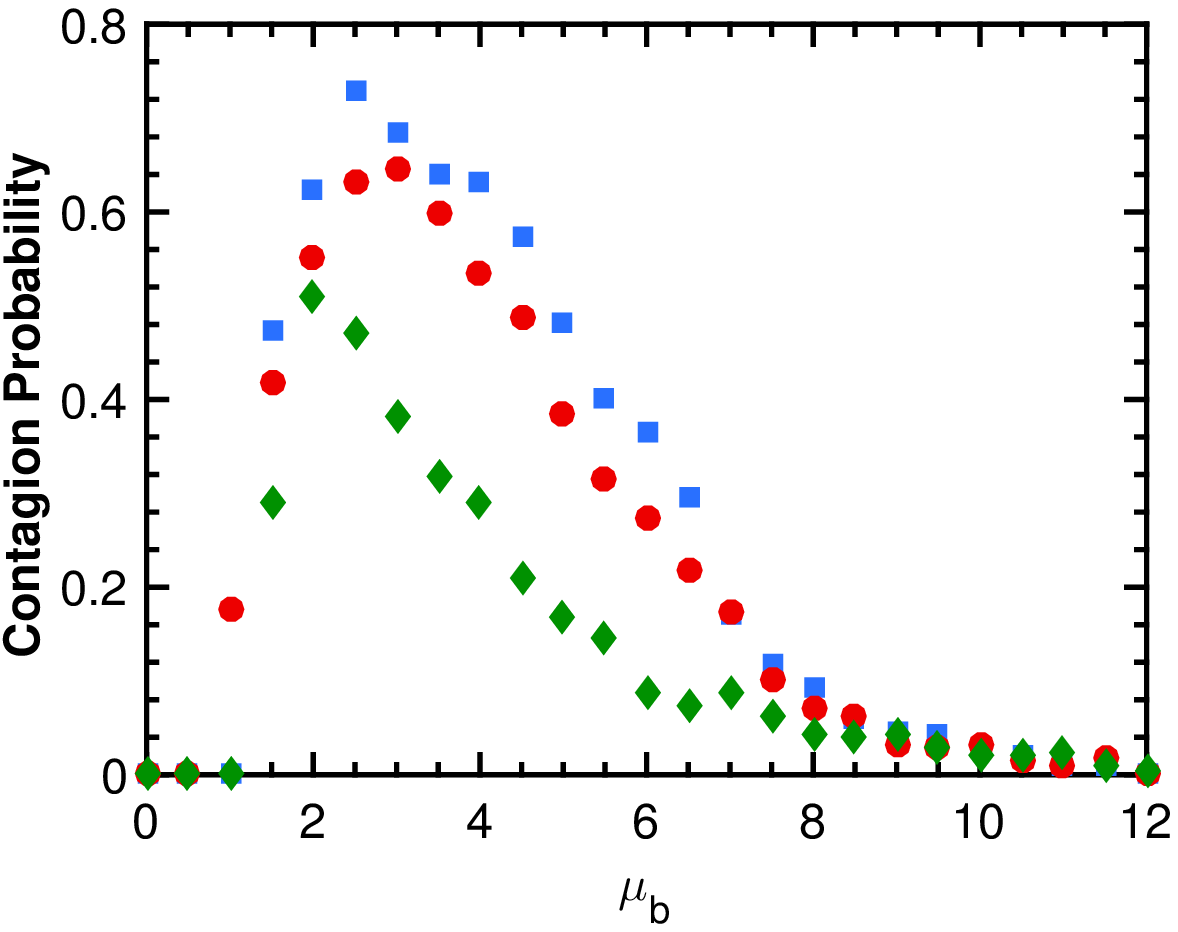}
			\caption{Asset shock}
			\label{f10sub2}
		\end{subfigure}
		\caption{Left Panel: Contagion probability as a function of $\mu_b$ for different network correlation configurations subject to the initial failure of a random bank. Blue squares: Uncorrelated network. Red circles: Assortative network. Green diamonds: disassortative network. The disassortative network gives the most stable configuration, while the assortative network results in the most unstable system. Right Panel: Contagion probability as a function of $\mu_b$ for different network correlation configurations. Again, the disassortative network gives the most stable configuration.}
		\label{fig10}
	\end{figure}
	
	In the left panel of \autoref{fig10}, we study the resilience of the system as a function of $\mu_b$ for the different network configurations for the case when a random bank fails. The right panel shows the same plot but for the case when a random asset is devalued. In both cases, we find that the disassortative network produces the most stable configuration. This is so because in a disassortative network, assets with high concentration are held by the most fragile banks (i.e. banks with low degrees). This implies that fire sales impact on the asset prices resulting from the default of any of these fragile banks would be minimal. However, in the assortative network, assets with low concentration degrees are held by these fragile banks, which implies that the fire sales resulting from their default would be much more severe thus leading to a wider unstable region. This result raises a question of whether it is possible to implement a structure of incentives that makes the bipartite network disassortative?
	
	\section{Impact of Leverage}
	\label{se:lev}
\begin{figure}[!htbp]
		\centering
		\includegraphics[width=.6\linewidth]{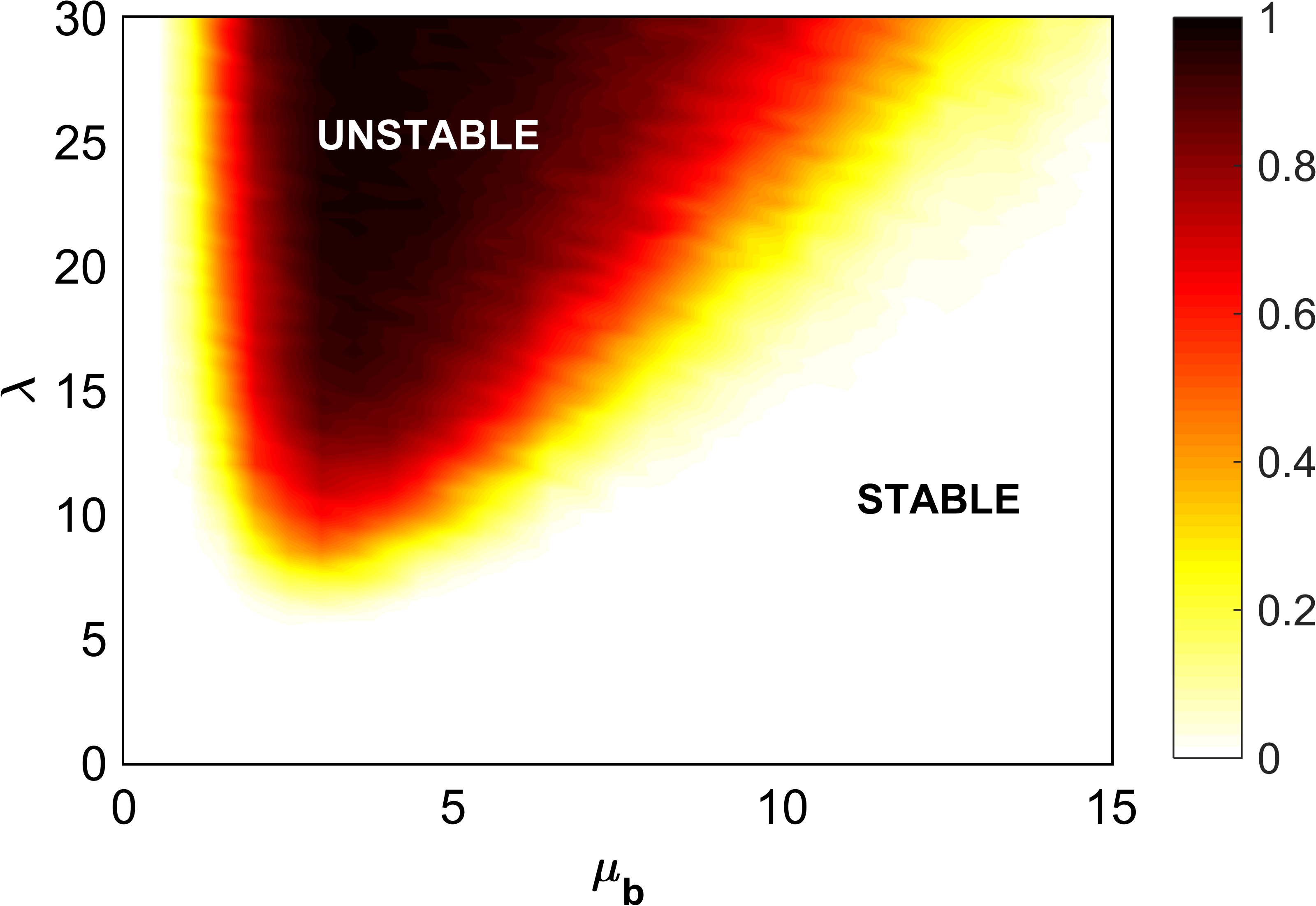}
		\caption{The non-white region refers to parameter values of $\lambda$ and $\mu_b$ that result in non-zero contagion probability. There is a critical leverage value below which the system is stable for any value of $\mu_b$.}
		\label{fig11}
	\end{figure}

	We now study the joint role of leverage (i.e. $\lambda$) and average degree (i.e. $\mu_b$) on the stability of our heterogeneous system. In \autoref{fig11}, we show that the existence of the critical leverage threshold for which contagion occurs with non-zero probability reported in \cite{Caccioli2014} for a homogeneous system is preserved when banks have heterogeneous degrees for each $\mu_b$ and that this threshold is increasing with $\mu_b$ irrespective of other prevailing conditions. This suggests that it may be possible for a financial regulator to permit higher leverage in the system by promoting an appropriate diversification strategy that achieves a particular value of $\mu_b$ which may not be individually optimal for the banks similar to the findings reported in \autocite{Tasca2014,Beale2011}.
	\section{Conclusion}
	\label{se:con}
	Previous studies on overlapping portfolios have relied on the assumption of homogeneity in the degrees and sizes of banks, however, empirical findings show that real financial networks deviate from this assumption \autocite{Guo2015,Braverman2014a,Boss2004,Marotta2015,DeMasi2012}. In particular, they provide evidence that bank degrees and sizes follow power law distributions. In our work, we generalised the model recently introduced in \cite{Caccioli2014} to account for these features. This approach makes it possible to study the systemic risk contribution of different types of banks with varying degrees and sizes. We found that separately introducing heterogeneity into the degrees and sizes of the banks widen the unstable region relative to the homogeneous case with respect to the initial failure of a random bank but not with respect to targeted shocks. In contrast, heterogeneity in asset concentrations makes the system more resilient to random shocks but not with respect to targeted shocks.
	
	Based on these intuitions, we proceeded to study possible capital policy models. Our findings suggest that a regulatory capital policy that assigns capital requirements to the most specialised banks performs better than random capital assignments when the network connectivity is high. However, focusing capital requirements on only the biggest bank does not appear to be effective relative to random assignments within the context of our model. Furthermore, we investigated the relevance of using diversification or size in building the capital based policies and find that the diversification based policy outperforms the size based policy with increasing network connectivity. 
	
	We then proposed a non-capital based policy that improves financial stability by introducing structural correlation into the bipartite network. Our results suggest that disassortative mixing (i.e. connecting the most specialised banks with the most concentrated assets) improves the resilience of the system. This can be understood from the fact that the fire sales impact of the specialised banks is significantly reduced due to the smaller quantity of traded shares relative to the entire volume of the assets. Finally, we studied the joint role of leverage and average degree on the stability of our heterogeneous system and found that the existence of a critical leverage beyond which contagion occurs with non-zero probability for each average degree reported in \cite{Caccioli2014} for a homogeneous system is preserved when banks have heterogeneous degree distribution. This finding further reinforces calls for policy makers to compensate for higher system risk induced by higher leverage by promoting an appropriate diversification strategy.
	
	In an ongoing work, we plan to break away from the mechanistic stress test models used in this paper and consider a more realistic agent based model in which the systemic risk from overlapping portfolios is endogenously created. This way we can implement measures to disincentive banks from structuring their portfolios in a manner that increases the fragility of the system. 
	
	\section*{Acknowledgements}
	This work was supported by the Nigerian Petroleum Trust Development Fund and the Quantitative \& Applied Spatial Economic Research Laboratory, University College London. We would like to thank Thomas Schroeder and J. Doyne Farmer for very helpful discussions and feedbacks at the initial stages of this project. OB appreciates Jadesola Onigbanjo for providing constructive evaluation of this paper. FC acknowledges support of the Economic and Social Research Council (ESRC) in funding the Systemic Risk Centre (ES/K002309/1).
	
	\printbibliography	
	
\end{document}